\documentclass[12pt]{article}
\usepackage{amssymb,amsmath,epsfig}

\begin{document}
\title{\bf Cosmological Evolution of Pilgrim Dark Energy}
\author{M. Sharif \thanks{msharif.math@pu.edu.pk} and M. Zubair
\thanks{mzubairkk@gmail.com}\\
Department of Mathematics, University of the Punjab,\\
Quaid-e-Azam Campus, Lahore-54590, Pakistan.}

\date{}

\maketitle

\begin{abstract}
We study pilgrim dark energy model by taking IR cut-offs as particle and
event horizons as well as conformal age of the universe. We derive evolution
equations for fractional energy density and equation of state parameters for
pilgrim dark energy. The phantom cosmic evolution is established in these
scenarios which is well supported by the cosmological parameters such as
deceleration parameter, statefinder parameters and phase space of
$\omega_\vartheta$ and $\omega'_\vartheta$. We conclude that the consistent
value of parameter $\mu$ is $\mu<0$ in accordance with the current Planck and
WMAP$9$ results.
\end{abstract}

\section{Introduction}

Over the past decade, the substantial progress in astronomical observations
indicate that our universe is presently going through the phase of
accelerated expansion. Observations of type Ia supernovae (SNeIa) (Perlmutter
et al. 1999; Riess et al. 2007), anisotropy measurement in current cosmic
microwave background (CMB) from WMAP (Spergel et al. 2004) and data of large
scale structure (LSS) from Salon Digital Sky Survey (SDSS) (Tegmark  et al.
2004) strongly endorse this manifestation. The mechanism behind the expanding
paradigm is usually assigned to exotic energy component with strong negative
pressure entitled as dark energy (DE). This may determine the ultimate future
of the universe but its cosmological origin and characteristics are still a
complicated story. The most likely theoretical campaigner of DE is the
cosmological constant $\Lambda$ with equation of state (EoS)
$\omega_\vartheta=-1$ (Weinberg 1989; Peebles and Ratra 2003). The model
comprising of $\Lambda$ and cold dark matter (CDM) dubbed as $\Lambda$CDM
model which suffers from fine tuning and cosmic coincidence puzzles. However,
the predictions of $\Lambda$CDM model appear to fit the current observational
data (Samushia and Ratra 2008; Jassal et al. 2008). Numerous candidates of DE
have been suggested in literature such as quintessence
($\omega_\vartheta>-1$), phantom ($\omega_\vartheta<-1$) violating the null
energy condition, quintom with $\omega_\vartheta$ evolving across $-1$,
K-essence, tachyon, ghost condensate, holographic DE (HDE) and so forth
(Arkani-Hamed et al. 2002; Armendariz-Picon et al. 2000; Caldwell 2002; Feng
et al. 2005; Hsu 2004; Li 2004; Steinhardt et al. 1999). Introducing new
ingredients of DE to the whole cosmic energy is one way to handle the issue
of cosmic acceleration. Another approach is the modification of Einstein
Lagrangian to get modified theories such as $f(R)$ (Sotiriou and Faraoni
2010), $f(R,T)$ (Harko et al. 2011; Sharif and Zubair 2012a, 2012b, 2013a,
2013b, 2013c) and $f(R,T,R_{\mu\nu}T^{\mu\nu})$ (Haghani et al. 2013; Sharif
and Zubair 2013d, 2013e) gravities, $T$ is the trace of the energy-momentum
tensor.

Cohen et al. (1999) set up a relation between ultraviolet (UV) and infrared
(IR) cut-offs due to the limit made by the formation of a BH. If
$\rho_\vartheta$ is the quantum zero-point energy density associated with UV
cut-off then entire energy in a sysytem of size $L$ should not exceed BH mass
of the same size so that $L^3\rho_\vartheta\leqslant{L}M_{p}^2$,
$M_p=1/\sqrt{8\pi{G}}$ is the reduced Planck mass. The largest IR cut-off
saturates the inequality and one gets the HDE density
\begin{equation*}
\rho_{\vartheta}=\frac{3c^2M^{2}_p}{L^{2}},
\end{equation*}
where $3c^2$ is a numerical constant. Several proposals have been suggested
for IR cut-off including Hubble, particle and event horizons as well as
conformal age of the universe, Ricci scalar and Granda Oliveros cut-off (Li
2004; Gao et al. 2009; Granda and Oliveros 2008, 2009; Wei and Cai 2008a).

Phantom form of DE $(\rho+p<0)$ possesses a peculiar feature of big rip, the
innumerous cosmic expansion within finite time. In such scenario, the energy
density grows quickly and disrupts all the large structures and bounded
objects. A question arises about the fate of BHs in the universe dominated by
phantom DE. One can say that repulsive force would be strong enough to avoid
the gravitational collapse and formation of BHs. Babichev et al. (2004)
explored phantom energy accretion of BH and found that its mass decreases
gradually. Some authors (Jamil and Qadir 2011; Sharif and Abbas 2011, 2012)
discussed this issue for different BH solutions. Gao et al. (2008) showed
that physical mass of BH may rather increase due to accretion of phantom
energy implying the violation of cosmic censorship conjecture.

Gonzalez and Guzman (2009) tested the accretion of phantom scalar field into
BH with different initial configurations and found that this mechanism can
reduce one half of the BH area. Sun (2009) studied dynamical equation of BH
mass in terms of cosmological parameters and obtained that BH mass reduces to
zero for the phantom dominated universe approaching to big rip. Recently, Wei
(2012) proposed a new model of DE named as pilgrim DE (PDE) based on the idea
that phantom DE is strong enough to avoid the formation of BH. He considered
Hubble horizon as an IR cut-off and developed constraints on PDE using the
latest cosmic observations. Sharif and Jawad (2013) analyzed the interacting
PDE models in terms of present day values of cosmographic parameters.

This paper explores the cosmological evolution of PDE for three cut-offs
namely particle horizon, event horizon and conformal age of the universe in
FRW universe. We follow the work of Li (2004) in HDE to explore the features
of non-interacting PDE for these cut-offs through fractional DE density
$\Omega_\vartheta$, EoS parameter $\omega_\vartheta$, statefinder diagnostic
parameters and $\omega_\vartheta-\omega'_\vartheta$ analysis. The paper has
the following format. In next section, we comprehensively present the
evolutionary paradigm of PDE. We conclude our results in the last section.

\section{Pilgrim Dark Energy}

The pilgrim dark energy is defined through the relation Wei (2012)
\begin{equation}\label{3}
\rho_{\vartheta}=3n^2M_p^{4-\mu}L^{-\mu}.
\end{equation}
The first Friedmann equation is given by
\begin{equation}\label{4}
3M_p^2H^2=\rho,
\end{equation}
where $\rho=\rho_M+\rho_{\vartheta}$ comprises of matter as well as DE
components and $H$ is the Hubble parameter. The matter energy density is
defined as $\rho_M=\rho_{M0}e^{-3x}~(x=\ln{a})$ from the matter energy
conservation equation. By setting the fractional energy densities of matter
and DE
\begin{eqnarray}\nonumber
\Omega_{M}=\frac{\rho_M}{\rho_{cri}},
\quad \Omega_{\vartheta}=\frac{\rho_{\vartheta}}{\rho_{cri}}, \quad \rho_{cri}=3M^{2}_pH^2,
\end{eqnarray}
Eq.(\ref{4}) can be cast to the form
\begin{equation}\label{5}
\Omega_m+\Omega_{\vartheta}=1,
\end{equation}
or
\begin{equation}\label{6}
H(x)=H_0\left(\frac{\Omega_{M0}e^{-3x}}
{1-\Omega_{\vartheta}}\right)^{1/2}.
\end{equation}
If $\Omega_{\vartheta}$ is known then one can determine the whole
expansion history $H(x)$. We discuss cosmological evolution for
different cut-offs such as particle and event horizons as well as
conformal age of the universe.

\subsection{Particle Horizon}

This horizon was initially used by Fischler and Susskind (1998) in
holographic cosmology.  Li (2004) discussed HDE by taking particle horizon as
an IR cut-off and found that it does not imply realistic cosmology with EoS
$\omega_\vartheta>-1/3$. The particle horizon is defined as
\begin{equation}\label{7}
L=R_{p}=a(t)\int^{t}_{0}{\frac{d\hat{t}}{a(\hat{t})}}
=a(t)\int^{a}_{0}{\frac{da'}{Ha'^{2}}}.
\end{equation}
Combining the definition of PDE (\ref{3}) and particle horizon
(\ref{7}), it follows that
\begin{equation}\label{8}
\int^{t}_{0}{\frac{d\hat{t}}{a(\hat{t})}}
=\int^{a}_{0}{\frac{da'}{Ha'^{2}}}=\frac{1}{a}\left(\frac{n^2M_p^{2-\mu}}{H^2
\Omega_{\vartheta}}\right)^{1/\mu}.
\end{equation}
Equation (\ref{5}) can be represented as
\begin{equation}\nonumber
\frac{1}{Ha}=\frac{\sqrt{a(1-\Omega_{\vartheta})}}{H_0\sqrt{\Omega_{m0}}}.
\end{equation}
Substituting this relation in Eq.(\ref{8}), we have
\begin{equation}\nonumber
\int^{a}_{0}{\sqrt{a(1-\Omega_{\vartheta})}}d\ln{a}=e^{(3/\mu-1)x}\left(
\frac{n^2M_p^{2-\mu}}{H_0^{2-\mu}\Omega_{m0}^{1-\mu/2}}\right)^{1/\mu}\left(\frac{1}
{\Omega_{\vartheta}}-1\right)^{1/\mu}.
\end{equation}
Differentiating it with respect to $x=\ln{a}$, it follows that
\begin{equation}\label{9}
\Omega'_{\vartheta}=\Omega_{\vartheta}(1-\Omega_{\vartheta})\left(3-\mu-
\frac{\mu}{C}(1-\Omega_{\vartheta})^{1/2-1/\mu}(\Omega_{\vartheta})^{1/\mu}
e^{(3/2-3/\mu)x}\right),
\end{equation}
where
$C=\left(\frac{n^2M_p^{2-\mu}}{H_0^{2-\mu}\Omega_{m0}^{1-\mu/2}}\right)^{1/\mu}$
and prime indicates derivative with respect to $x=\ln{a}$. This
result can explain the cosmic evolution according to PDE with
particle horizon.

One can exactly solve the above equation to represent the behavior of PDE (Li
2004). The corresponding EoS parameter can be set by using the energy
conservation equation as
\begin{equation}\label{10}
\omega_\vartheta=-1-\frac{1}{3}\frac{d\ln{\rho_\vartheta}}{dx}=-1+\frac{\mu}{3}
\left(1-\frac{1}{H}\left(\frac{H^2\Omega_{\vartheta}}{n^2M_p^{2-\mu}}\right)^{1/\mu}\right).
\end{equation}
If we set $\mu=2$, Eqs.(\ref{9}) and (\ref{10}) reproduce the corresponding
results in HDE with particle horizon as an IR cut-off (Li 2004). For HDE, EoS
is $\omega_\vartheta=-1+\frac{2}{3n}\sqrt{\Omega_\vartheta}$ indicating that
when $\Omega_\vartheta\rightarrow1$ in the future,
$\omega_\vartheta=-1+\frac{2}{3n}>\frac{-1}{3}$ which appears to be
inconsistent with the accelerating phase. In case of PDE, we have dependence
on parameter $\mu$ resulting in extra degrees of freedom. As $n^2$ is
involved in Eqs.(\ref{9}) and (\ref{10}), so the expansion history is
independent of signature of $n$. Solving Eq.(\ref{9}) with the initial
condition $\Omega_{\vartheta0}=1-\Omega_{m0}$ and using in Eq.(\ref{10}), the
evolution of PDE is shown in Figures \textbf{1-4}. Here, present day values
of $\Omega_m$ and $H$ are defined from recent Planck results as
$\Omega_{m0}=0.315$ and $H_0=67.3$. The Planck and WMAP$9$ observations set
the constraints for EoS of DE $\omega_\vartheta$ as
$\omega_\vartheta=-1.13^{+0.13}_{-0.10}$ and $-1.71<\omega_\vartheta<-0.34$
respectively (Ade et al. 2013)
\begin{figure} \centering
\epsfig{file=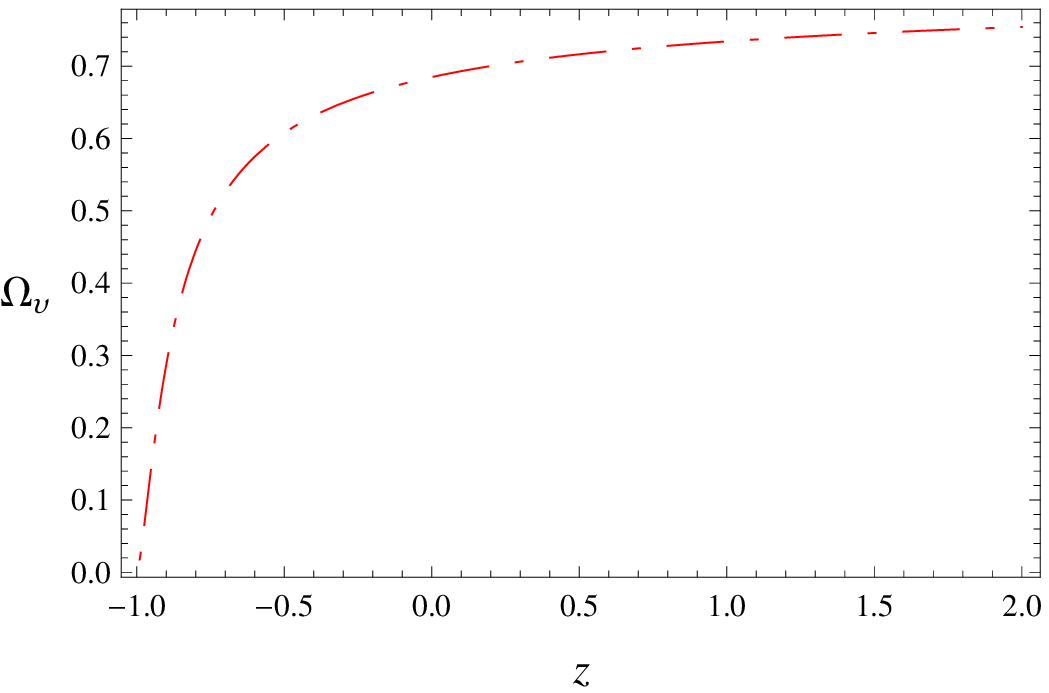, width=.495\linewidth, height=2.2in}
\epsfig{file=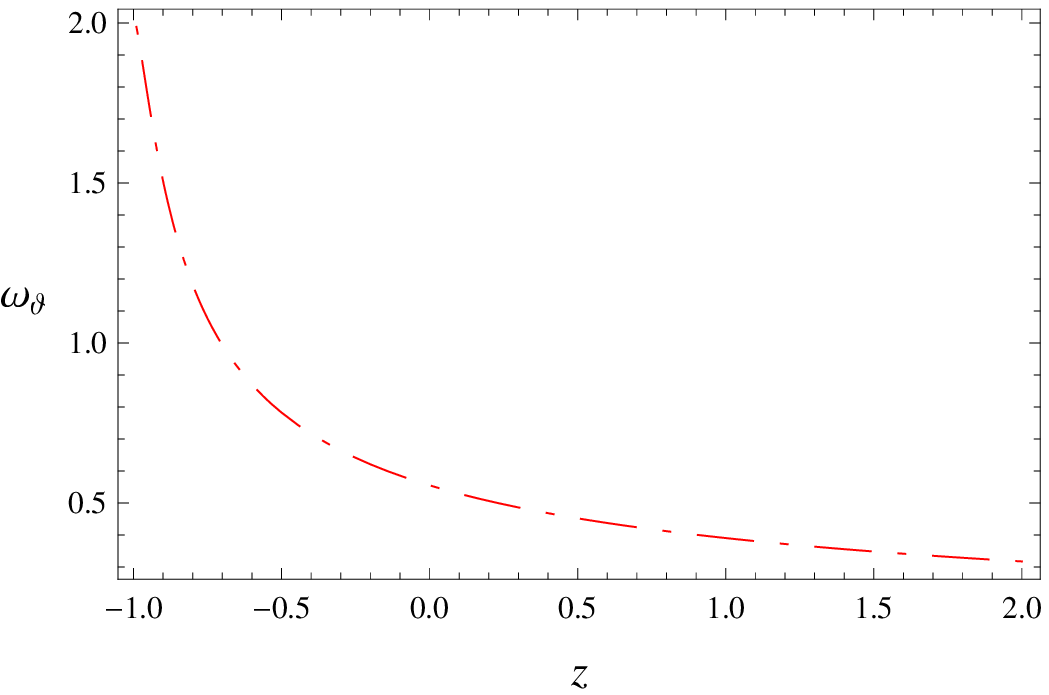, width=.495\linewidth, height=2.2in}
\caption{Evolution of $\Omega_\vartheta$ and $\omega_\vartheta$ for
PDE with particle horizon. Here we set $\mu=3$ and $n=2$. It clearly
shows that such choice is no more realistic resulting matter
dominated regime in the future evolution.}
\end{figure}

For $\mu\geqslant3$, we have purely matted dominated phase of the universe as
shown in Figure \textbf{1}. This choice should be neglected in search of some
observationally consistent models. If $\mu<0$, the evolution of
$\omega_\vartheta$ and $\Omega_\vartheta$ are shown in Figure \textbf{2}
which indicates that $\omega_\vartheta$ is always in phantom region and never
intersects the phantom divide line ($\omega_\vartheta=-1$) in entire cosmic
evolution. This behavior is similar to the case of PDE with Hubble horizon in
which $\omega_\vartheta<-1$ in whole cosmic history (Wei 2012). For the
Hubble horizon, $\omega_\vartheta$ asymptotically goes to $-1$, \emph{i.e.},
it represents de Sitter phase in late times whereas in our case for PDE with
particle horizon, it ends up with phantom phase.
\begin{figure}
\centering \epsfig{file=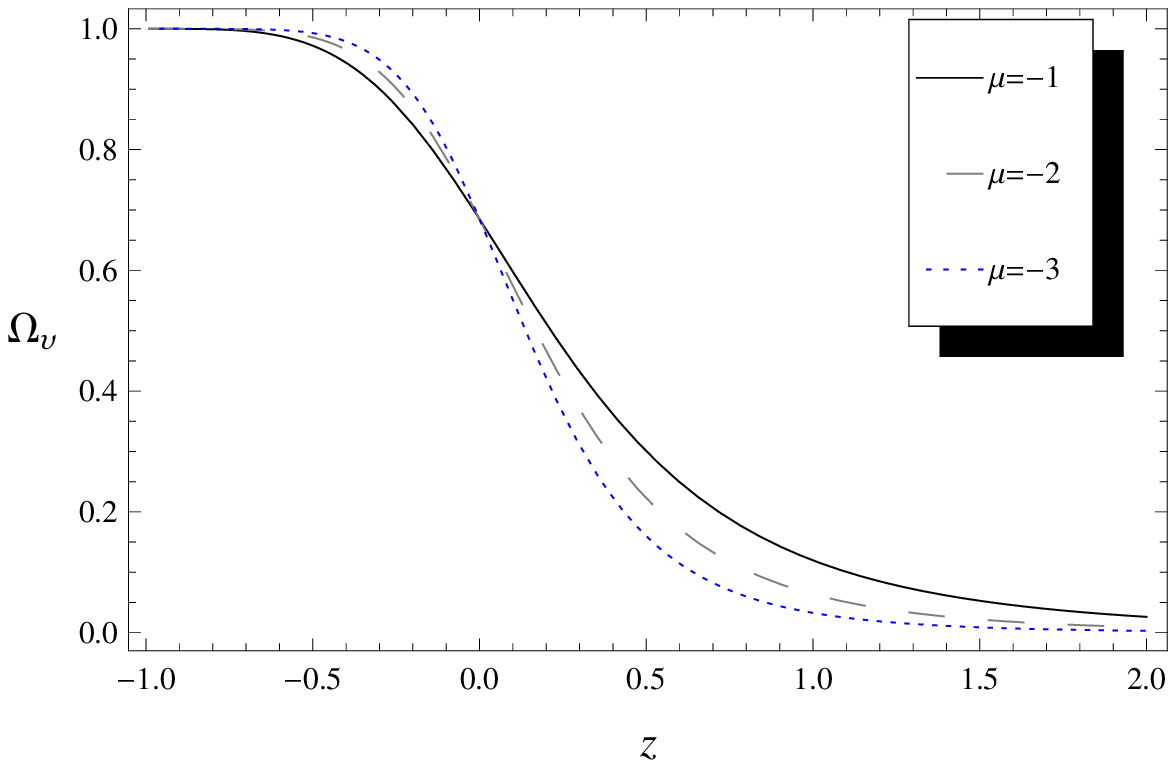, width=.495\linewidth,
height=2.2in} \epsfig{file=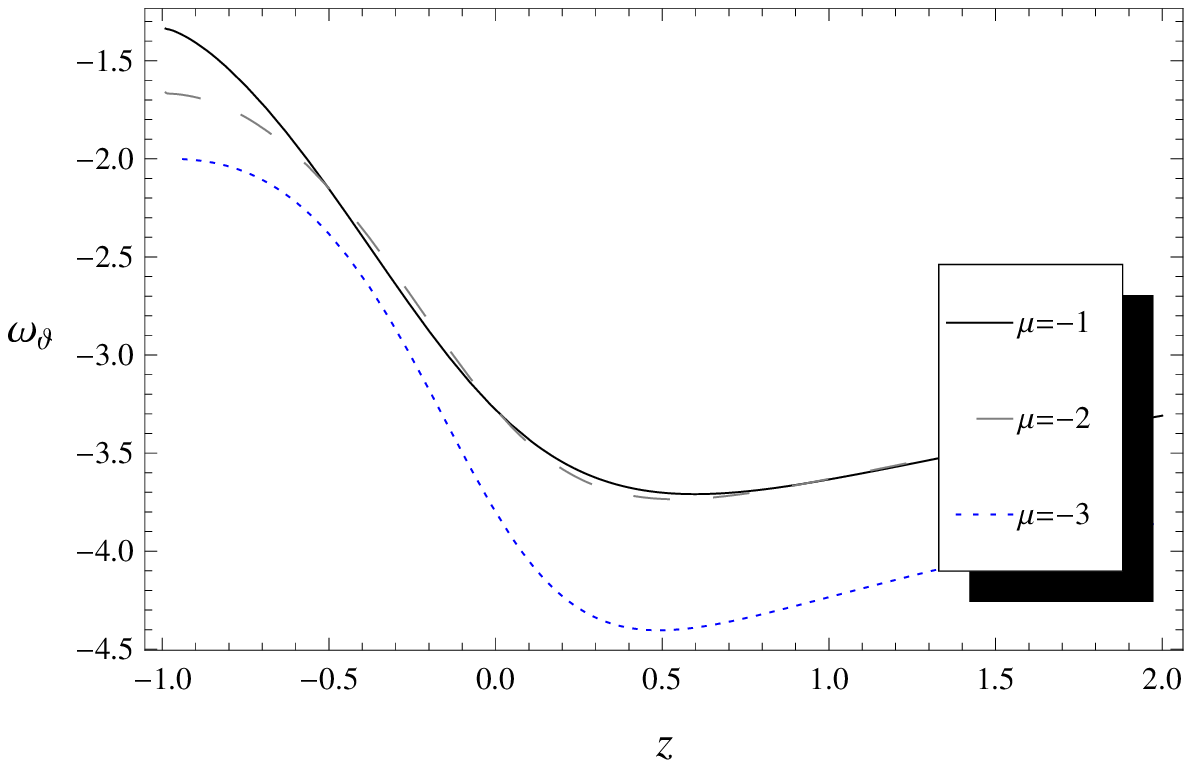, width=.495\linewidth,
height=2.2in} \caption{Evolution of $\Omega_\vartheta$ and $\omega_\vartheta$
for PDE with particle horizon ($\mu\leqslant-1$).}
\end{figure}

The deceleration parameter is defined in terms of $\omega_\vartheta$ and
$\Omega_\vartheta$ as
\begin{equation}\nonumber
q=-\frac{a\ddot{a}}{\dot{a}^2}=\frac{1}{2}(1+3\omega_\vartheta\Omega_\vartheta),
\end{equation}
which is a handy tool to explain the transition from decelerated phase to
accelerating regime. We plot $q$ versus $z$ and show the transition from
decelerated phase to accelerated era. The universe entered in accelerated era
in recent past and it will finish with $q<-1$ representing the phantom
evolution. The change of signature in $q$ depends upon the values of $\mu$
and the era of accelerated expansion begins earlier for large values of
$\mu$. Differentiating Eq.(\ref{10}) with respect to $x=\ln{a}$, we get
\begin{equation}\label{11}
\omega'_\vartheta=\left(1-\frac{3}{\mu}(1+\omega_\vartheta)\right)\left(-1+\frac{\mu}{2}
+\frac{1}{3}\frac{d}{dx}\ln{\Omega_\vartheta}\right).
\end{equation}

Caldwell and Linder (2005) discussed the quintessence feature of DE candidate
and analyzed its representation in $\omega_\vartheta-\omega_\vartheta'$
plane. They established the limits of quintessence model in phase space of
$\omega_\vartheta$ and $\omega'_\vartheta$ and pointed out two regions of
this plane namely thawing ($\omega'_\vartheta>0$ with $\omega_\vartheta<0$)
and freezing ($\omega'_\vartheta<0$ with $\omega_\vartheta<0$). It is
remarked that cosmic expansion is accelerated in freezing region when
compared with thawing region. This approach has been applied in different
settings by considering various forms of DE such as quintessence, phantom and
quintom models (Chiba 2006; Gao et al. 2006; Scherrer 2006). In (Sharif and
Zubair 2013f), we have also discussed the phase space of $\omega_\vartheta$
and $\omega'_\vartheta$ for new HDE which exhibits $\Lambda$CDM model
($\omega_\vartheta=-1$ and $\omega'_\vartheta=0$) in future evolution. The
evolution of $\omega'_\vartheta$ in $\omega_\vartheta-\omega'_\vartheta$
plane is shown in the right panel of Figure \textbf{3}. This represents the
freezing region for noninteracting PDE with particle horizon which favors the
phantom evolution in this format of DE.
\begin{figure}
\centering \epsfig{file=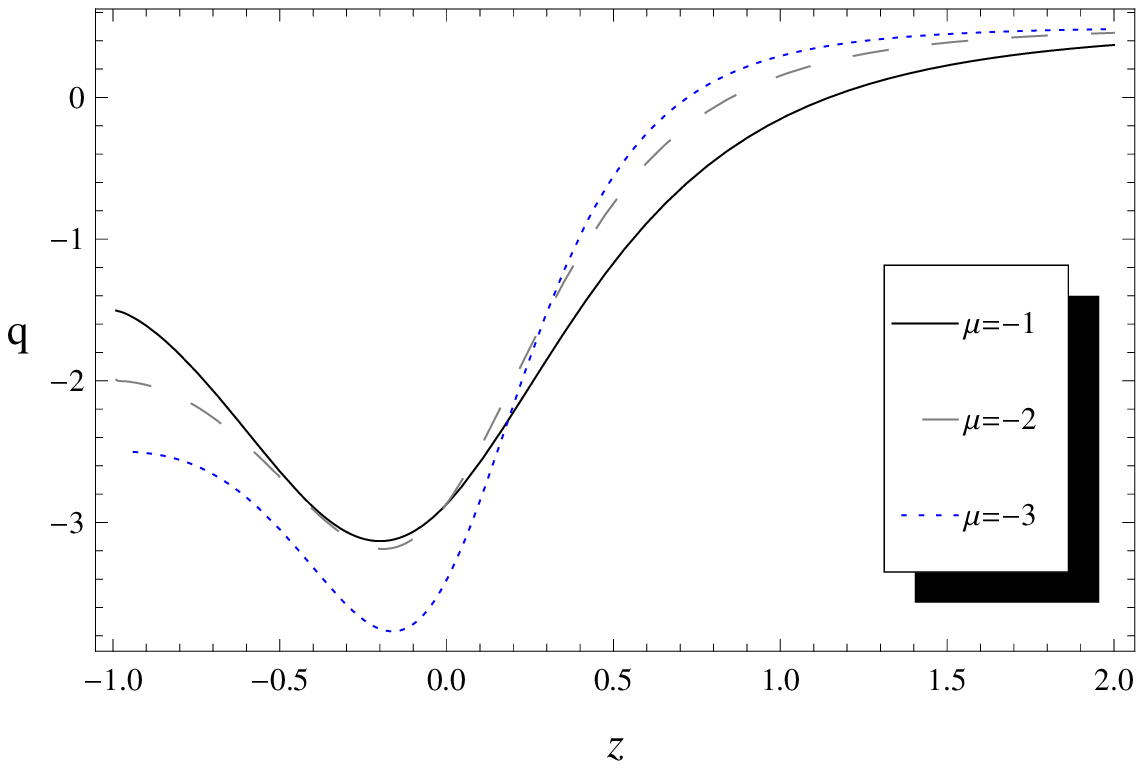, width=.495\linewidth,
height=2.2in} \epsfig{file=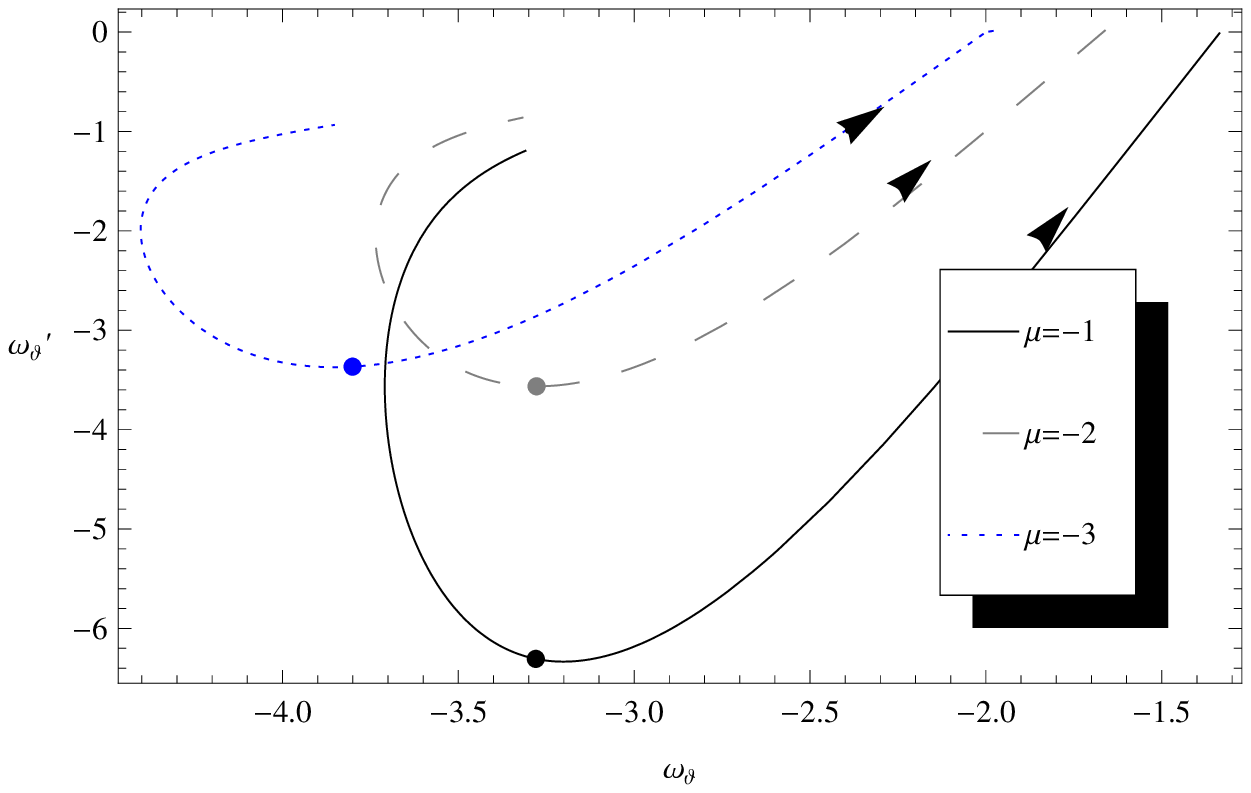, width=.495\linewidth,
height=2.2in} \caption{Evolution of $q$ and
$\omega_\vartheta-\omega'_\vartheta$ plane for PDE with particle
horizon for $\mu\leqslant-1$. A sign flip of $q$ indicates
transition to accelerated expansion and
$\omega_\vartheta-\omega'_\vartheta$ plane indicates the freezing
region. The dots represent present day values of
parameters.}
\end{figure}

In what follows, we examine PDE with particle horizon using the statefinder
diagnostic. The pair $\{r,s\}$ of statefinder diagnostic parameters is
defined as (Sahni et al. 2003)
\begin{eqnarray}\label{12}
r=\frac{\dddot{a}}{aH^3},\quad s=\frac{(r-1)}{3(q-1/2)},
\end{eqnarray}
$r$ is also named as jerk parameter. Statefinder diagnostic depends on the
scale factor, its derivatives $\dddot{a}$ and deceleration parameter $q$ to
differentiate the cosmic expansion on geometric grounds. Cosmological models
can be differentiated on the basis of statefinder diagnostic as it shows
distinct trajectories corresponding to specific models. For $\Lambda$CDM
model, the statefinder parameters are fixed as $(r, s)=(1,0)$ and in case of
CDM regime these correspond to $(r, s)=(1,1)$. In $r-s$ plane, the
trajectories for quintessence and phantom lie in the range $(s>0,~r<1)$
whereas for chaplygin gas these correspond to $(s<0,~r>1)$. The statefinder
diagnostic parameters can be represented in terms of $\omega_\vartheta$ and
$\Omega_\vartheta$ as
\begin{eqnarray}\label{13}
r=1-\frac{3}{2}\Omega_{\vartheta}\left[\omega_{\vartheta}'
-3\omega_{\vartheta}(1+\omega_{\vartheta})\right], \quad
s=\frac{-1}{3\omega_{\vartheta}}\left[\omega_{\vartheta}'
-3\omega_{\vartheta}(1+\omega_{\vartheta})\right].
\end{eqnarray}
\begin{figure}
\centering \epsfig{file=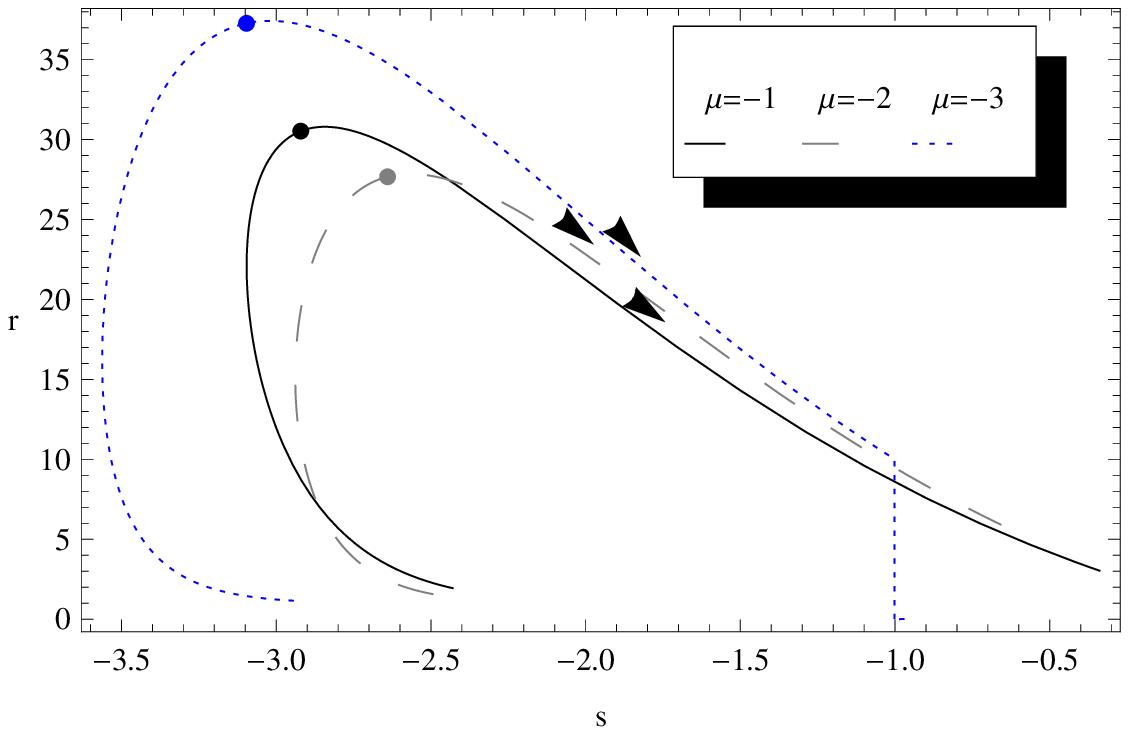, width=.495\linewidth,
height=2.2in} \epsfig{file=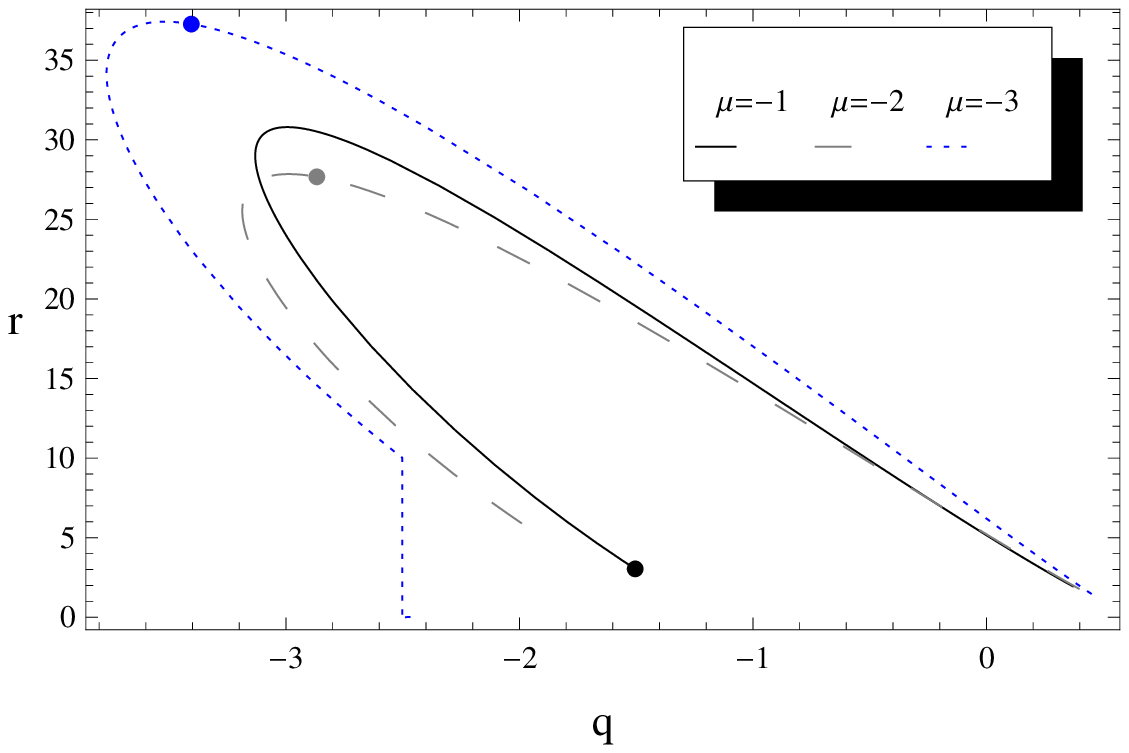, width=.495\linewidth,
height=2.2in} \caption{Evolution of statefinder diagnostic parameters for PDE
with particle horizon ($\mu\leqslant-1$).}
\end{figure}
Figure \textbf{4} shows the evolution trajectories of PDE with particle
horizon in $r-s$ and $q-r$ planes. In the left panel, the evolution
trajectories favor the chaplygin gas model with $s<0$ and $r>1$. We also plot
the evolution trajectories of the deceleration parameter in $q-r$ plane. Our
results are consistent with (Wu and Yu 2005, 2006) where authors performed
the statefinder diagnostic for the phantom and quintom DE model. This shows
that the non-interacting PDE with particle horizon favors the phantom regime
which is the basic idea of this candidate. Thus, for the realistic model of
PDE, one needs to set $\mu<0$ and this choice is well supported by the
results of Planck and WMAP$9$ observations (Ade et al. 2013; Bennet 2012).
Here we take $\mu\leqslant-1$ for which PDE implies that
$\omega_\vartheta<-1$ which is supported by other cosmographic parameters as
shown in Figures \textbf{3} and \textbf{4}.

\subsection{Event Horizon}
%
The IR cutoff $L$ (event horizon) is defined as
\begin{equation}\label{14}
L=R_{\hat{E}}=a(t)\int^{\infty}_{t}{\frac{d\hat{t}}{a(\hat{t})}}
=a(t)\int^{\infty}_{a}{\frac{da'}{Ha'^{2}}} .
\end{equation}
Employing the definition of PDE (\ref{3}) and event horizon (\ref{14}), we
obtain the dynamical equation of fractional density of DE as
\begin{equation}\label{15}
\Omega'_{\vartheta}=\Omega_{\vartheta}(1-\Omega_{\vartheta})\left(3-\mu+
\frac{\mu}{C}(1-\Omega_{\vartheta})^{1/2-1/\mu}(\Omega_{\vartheta})^{1/\mu}
e^{(3/2-3/\mu)x}\right).
\end{equation}
The time derivative of PDE with event horizon as an IR cut-off is
\begin{equation}\nonumber
\dot{\rho}_\vartheta=-\mu\rho_\vartheta\frac{\dot{L}}{L}, \quad \dot{L}=HL-1.
\end{equation}
Using the energy conservation equation of DE, we obtain
\begin{equation}\label{16}
\omega_\vartheta=-1+\frac{\mu}{3}\left(1-\frac{1}{H}\left(\frac{H^2\Omega_{\vartheta}}
{n^2M_p^{2-\mu}}\right)^{1/\mu}\right).
\end{equation}
One can reproduce the corresponding results in HDE with event horizon for
$\mu=2$. In case of HDE with event horizon, the EoS parameter is
$\omega_{\vartheta}=-\frac{1}{3}\left(1+\frac{2\sqrt{\Omega_{\vartheta}}}{n}\right)$
which can result in three significant eras of cosmic expansion. If the
universe is dominated by DE components, \emph{i.e.},
$\Omega_{\vartheta}\longrightarrow1$ in the future then for
$n>1,~\omega_{\vartheta}$ is always greater than $-1$ which depicts
quintessence era so that the universe escapes from getting in de Sitter and
big rip phases. For $n=1$, the universe enters the de Sitter era in future
evolution and $n<1$ represents phantom phase where the universe behaves as
quintom model of DE as $\omega_\vartheta$ intersects the cosmological
constant line. The value of $n$ plays a vital role in deciding the
evolutionary features of HDE and ultimate fate of the universe.
\begin{figure}
\centering \epsfig{file=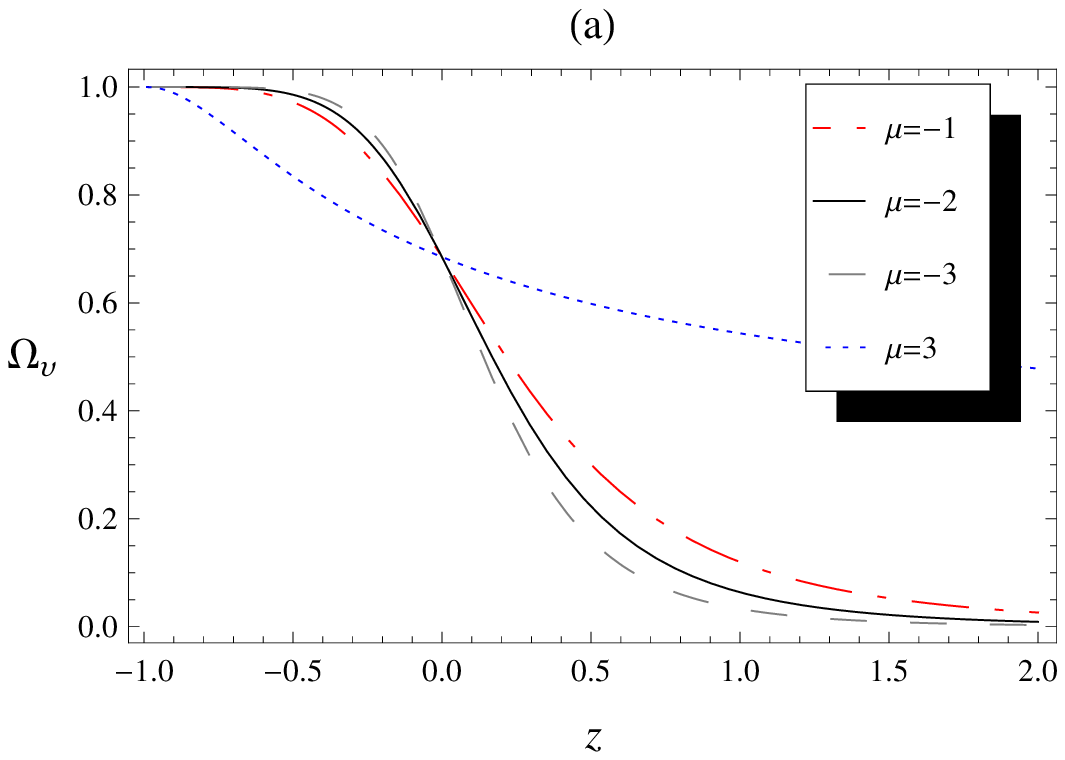, width=.495\linewidth,
height=2.2in} \epsfig{file=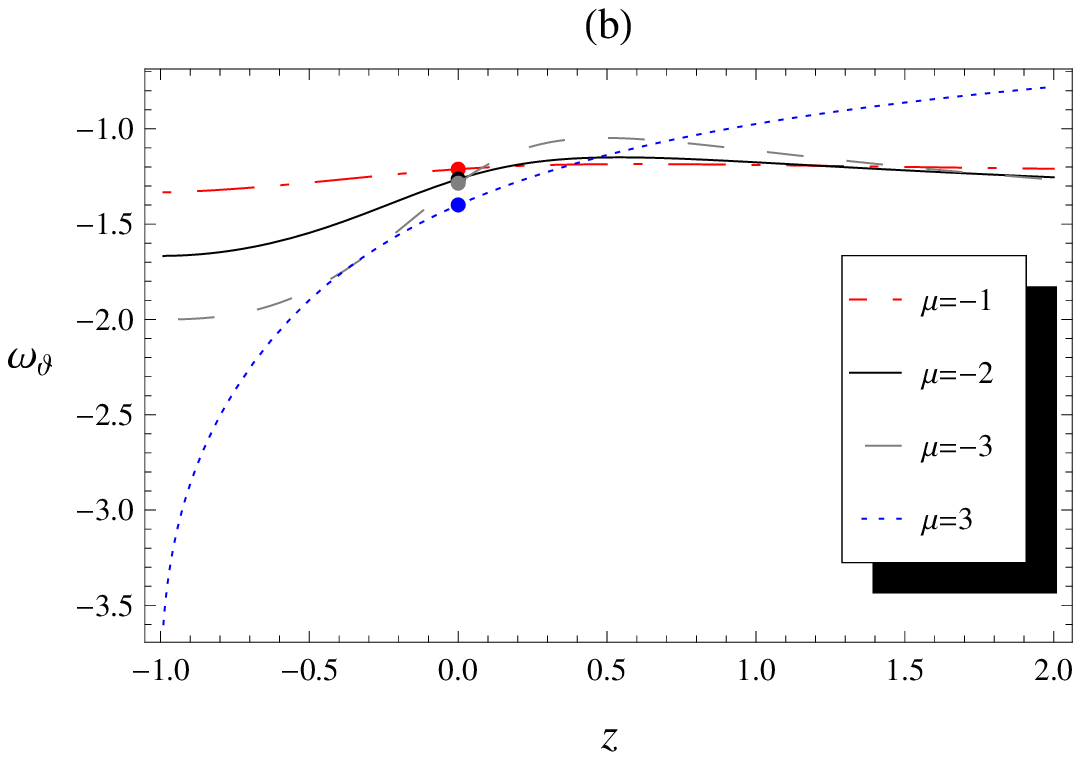, width=.495\linewidth,
height=2.2in} \caption{Evolution of $\Omega_\vartheta$ and
$\omega_\vartheta$ versus $z$ for PDE with event horizon for $n=0.5$
and different values of $\mu$. The dots indicate the present day values.}
\end{figure}

For PDE, the role of $\mu$ is more crucial as compared to that of $n$. In
this setting, we are mainly concerned with the choice $\mu<0$ but for PDE
with event horizon, one can also set $\mu=3$. The evolution trajectories of
EoS and fractional energy density of PDE are shown in Figure \textbf{5}. For
$\mu\leqslant-1$, $\Omega_\vartheta$ approaches to $1$ as $z\rightarrow-1$
showing that DE dominates in later times of the universe. The EoS parameter
$\omega_\vartheta$ is in the phantom regime (Figure \textbf{5(b)}) and the
present day values of  $\omega_\vartheta$ are consistent with the Planck
results showing $\omega_\vartheta=-1.13^{+0.13}_{-0.10}$ (Bennet 2012). In
case of $\mu>2$, we show the evolution trajectories for $\mu=3$ which does
favor the phantom feature of DE. For $\mu=3$, the curves in Figure \textbf{5}
show somewhat distinct behavior where $\omega_\vartheta<-3.5$ as
$z\rightarrow-1$. We neglect the values of $\mu>3$ because these values do
not imply realistic results. The evolution of $q$ versus $z$ is represented
in Figure \textbf{6(a)} which shows that the universe entered into the
accelerating phase in the recent past and it would switch over to $q<-1$
indicating the phantom paradigm.
\begin{figure}
\centering \epsfig{file=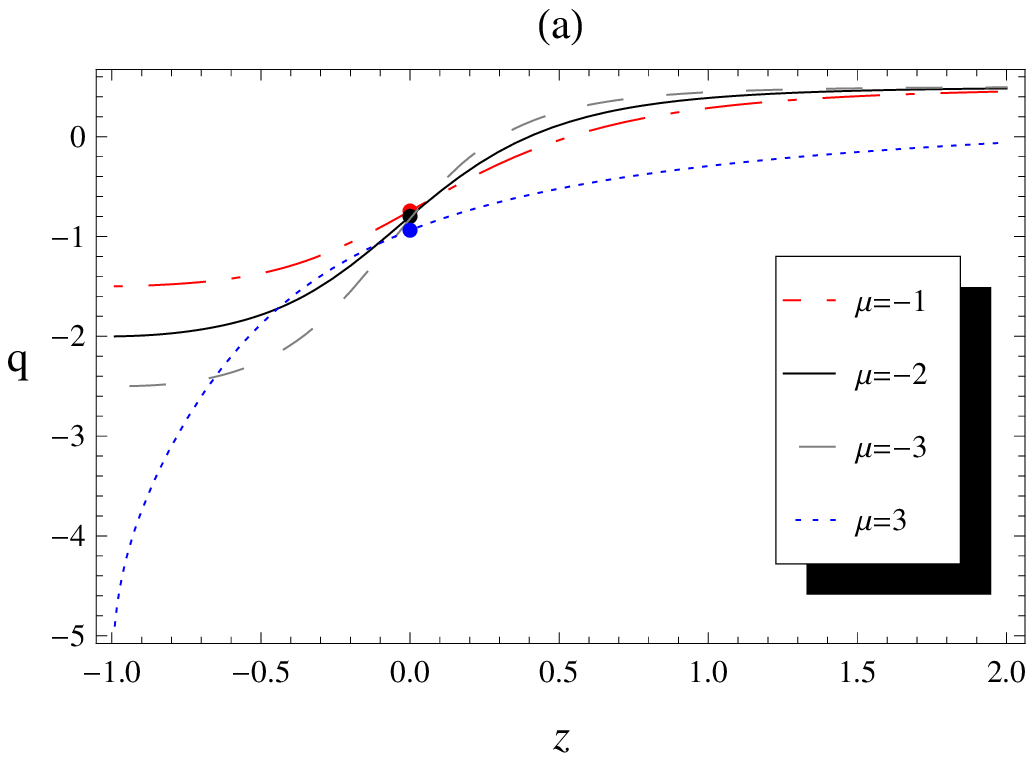, width=.495\linewidth,
height=2.2in} \epsfig{file=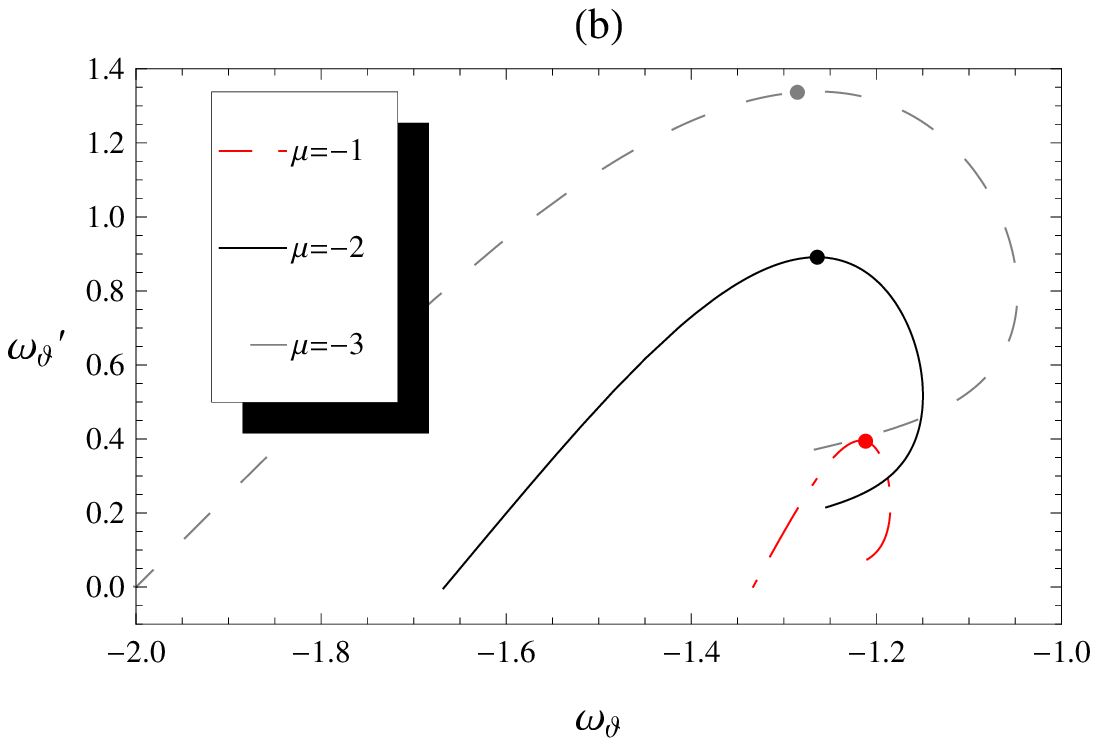, width=.495\linewidth,
height=2.2in} \caption{Plot \textbf{6(a)} shows the variation of $q$
versus $z$ and \textbf{6(b)} represents evolution trajectories of
$\omega_\vartheta-\omega'_\vartheta$ for non-interacting PDE with
event horizon corresponding to $n=0.5$ and different values of
$\mu$. The dots indicate the present day values of $q$ and
$\{\omega_\vartheta, \omega'_\vartheta\}$.}
\end{figure}
\begin{figure}
\centering \epsfig{file=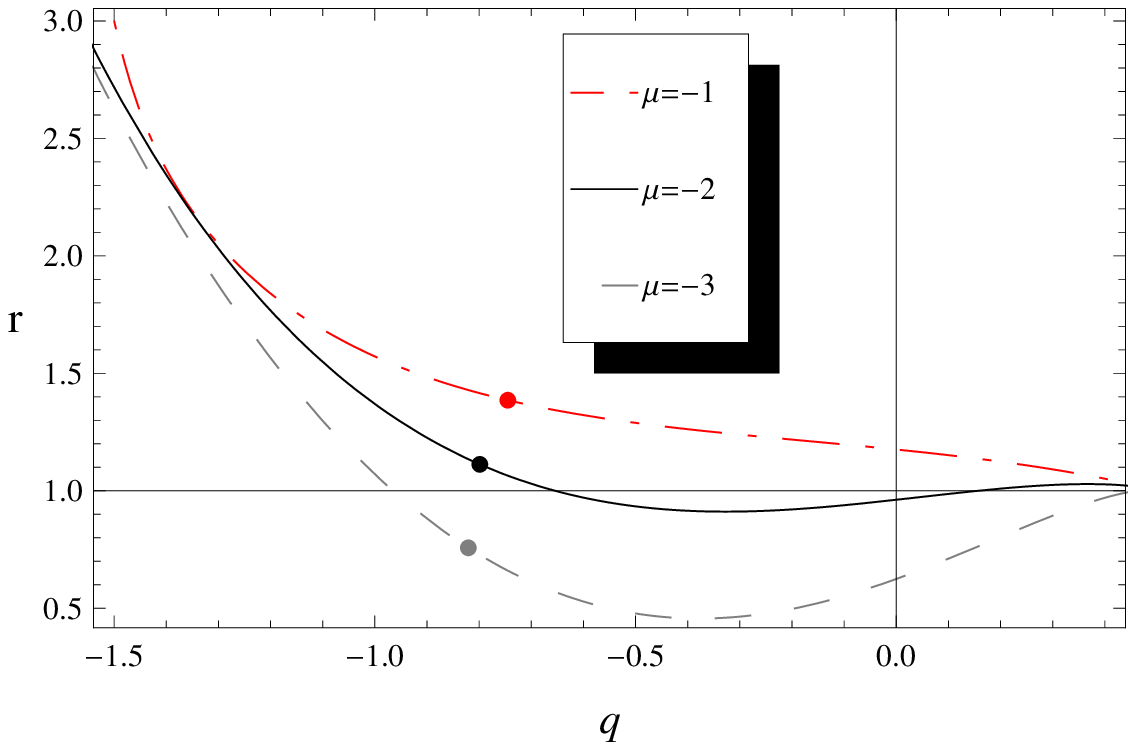, width=.495\linewidth,
height=2in}\epsfig{file=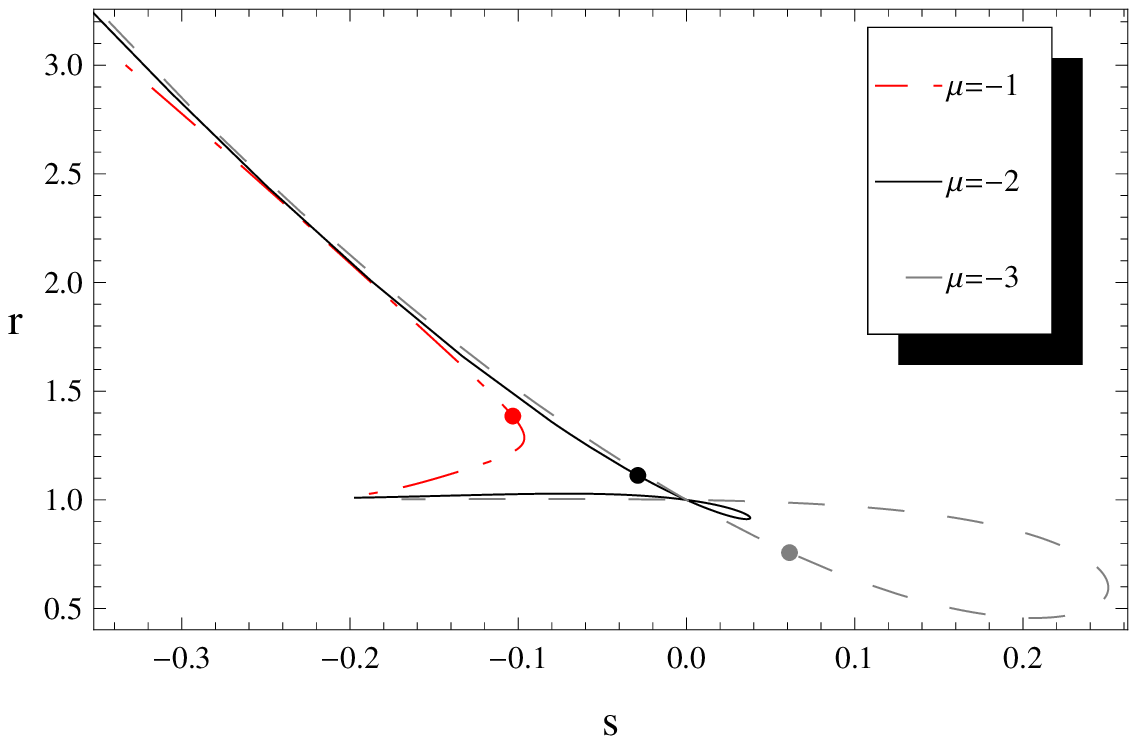, width=.495\linewidth,
height=2in}\caption{The statefinder analysis of PDE with event
horizon in $q-r$ and $r-s$ planes.}
\end{figure}

Taking derivative of Eq.(\ref{16}) with respect to $x$, we get
\begin{equation}\label{17}
\omega'_\vartheta=\left(1-\frac{3}{\mu}(1+\omega_\vartheta)\right)\left(1-\frac{\mu}{2}
-\frac{1}{3}\frac{d}{dx}\ln{\Omega_\vartheta}\right).
\end{equation}
The phase plane of $\omega_\vartheta$ and $\omega'_\vartheta$ for
non-interacting PDE with event horizon is shown in Figure \textbf{6(b)}. The
plane $\omega_\vartheta-\omega'_\vartheta$ represents the thawing region in
the evolution of PDE with event horizon. In the limit of future evolution
$z\rightarrow-1$, $\omega'_\vartheta\rightarrow0$ and $\omega_\vartheta>-1$.
Figure \textbf{7} shows statefinder analysis in $q-r$ plane for the choice
$\mu\leqslant-1$. The evolution trajectories in $q-r$ plane for PDE with
event horizon start from $(q<0.5, r<1)$ and end up with $(q<-1, r>1)$.

\subsection{Conformal Age of the Universe}

The two time scales, age of the universe and conformal time have been
suggested in literature corresponding to agegraphic DE (ADE) (Cai 2007) and
new agegraphic DE (NADE) (Wei and Cai 2008a). These models can derive the
cosmic expansion consistent with the recent observational data (Wei and Cai
2008b) which can resolve the causality problem. However, it is pointed out
that ADE model is classically unstable and NADE is no better than HDE in
explaining the DE dominated universe. The NADE has been studied to address
various cosmological issues in Einstein and modified gravities (Karami 2010;
Jamil and Saridakis 2010; Sheykhi 2010). The conformal age of the universe is
defined as
\begin{equation}\label{18}
\eta=\int{\frac{dt}{a(t)}} =\int{\frac{da}{Ha^{2}}}.
\end{equation}
Using Eqs.(\ref{3}) and (\ref{18}), the rate of change of fractional
DE density is
\begin{equation}\label{19}
\Omega'_{\vartheta}=\Omega_{\vartheta}(1-\Omega_{\vartheta})\left(3-\mu-
\frac{\mu}{C}(1-\Omega_{\vartheta})^{1/2-1/\mu}(\Omega_{\vartheta})^{1/\mu}
e^{(1/2-3/\mu)x}\right).
\end{equation}
The time derivative of $\rho_\vartheta$ is obtained as
\begin{equation}\nonumber
\dot{\rho}_\vartheta=-\frac{\mu\rho_\vartheta}{a}\left(\frac{H^2\Omega_{\vartheta}}
{n^2M_p^{2-\mu}}\right)^{1/\mu}.
\end{equation}
The corresponding EoS parameter is
\begin{equation}\label{20}
\omega_\vartheta=-1+\frac{\mu}{3aH}\left(\frac{H^2\Omega_{\vartheta}}
{n^2M_p^{2-\mu}}\right)^{1/\mu}.
\end{equation}

To demonstrate the evolution trajectories for NADE version of PDE, we include
some facts about non-interacting NADE model. Wei and Cai (2008b) showed that
coincidence problem can be resolved for NADE if one chooses the value of
parameter $n$ nearly unity. The NADE is constrained from the observational
data of SNeIa, CMB and LSS which implies the best fit value of
$n=2.76^{+0.111}_{-0.109}$ (with $1\sigma$ uncertainty). They found that EoS
parameter for NADE approaches to $-1$ in later times regardless of the value
of $n$. Zhang et al. (2013) showed WMAP 7-years observations set appropriate
measure of $n$ as $n=2.673^{+0.053+0.127+0.199}_{-0.077-0.151-0.222}$. In
previous study (Sharif and Zubair 2013f), we have reconstructed $f(R)$
gravity corresponding to NADE and set the values of $n$ as
$n=2.3,~2.8,~3.3,~3.8$. It has been shown that these parametric values
support the $\Lambda$CDM model in future evolution.

For PDE with conformal time scale, we are mainly concerned to explore the
behavior of parameter $\mu$ in determining the evolution for conformal time.
Initially, we set $\mu<0$ and found that $\mu=-1$ supports the cosmological
constant regime in future evolution as shown in Figures \textbf{8} and
\textbf{9}. Figure \textbf{8(a)} shows that $\Omega_\vartheta\rightarrow1$ in
the future evolution so that our universe is dominated by DE.
$\omega_\vartheta$ is always less than $-1$ in the whole cosmic history and
it will asymptotically approach to $-1$ in the future evolution of the
universe. The present behavior of EoS parameter favors the phantom DE
consistent with Planck and WMAP$9$ results. Therefore, evolution of the
universe will end up with $\Lambda$CDM model rather than big rip. Such
behavior is identical to that suggested by Wei and Cai (2008b) for PDE with
Hubble horizon. In Figure \textbf{9(a)}, the evolution of deceleration
parameter shows the bouncing behavior of the universe which entered in the
expansion phase in recent past and it would end up in a de Sitter phase.
\begin{figure}
\centering \epsfig{file=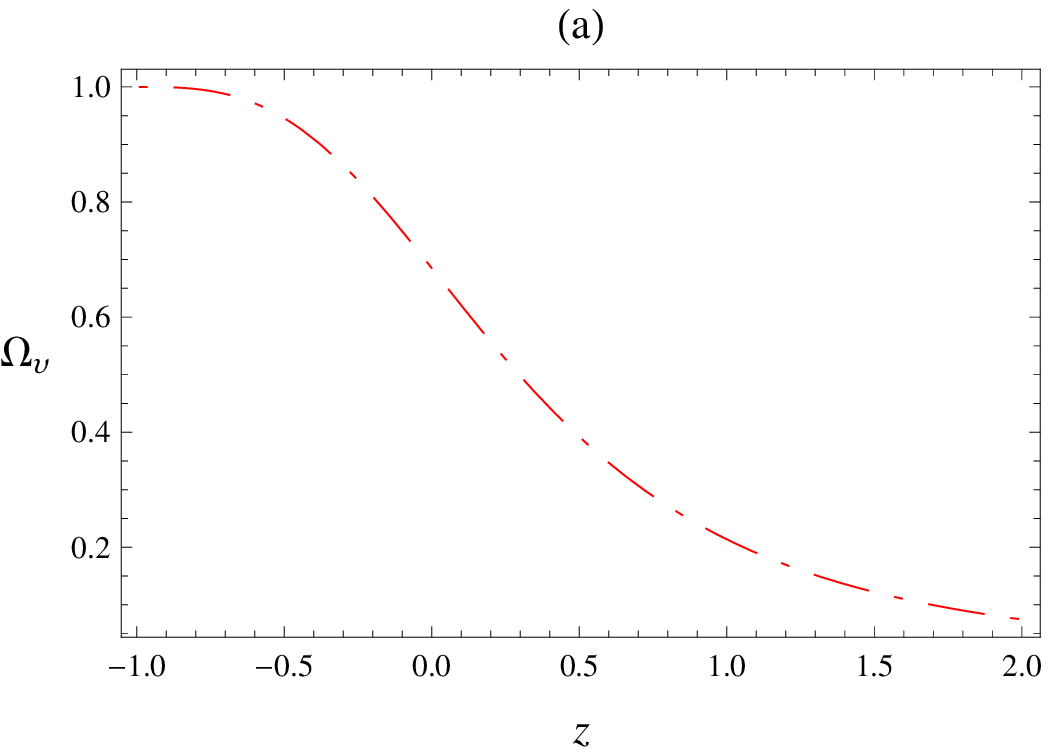, width=.495\linewidth,
height=2.2in} \epsfig{file=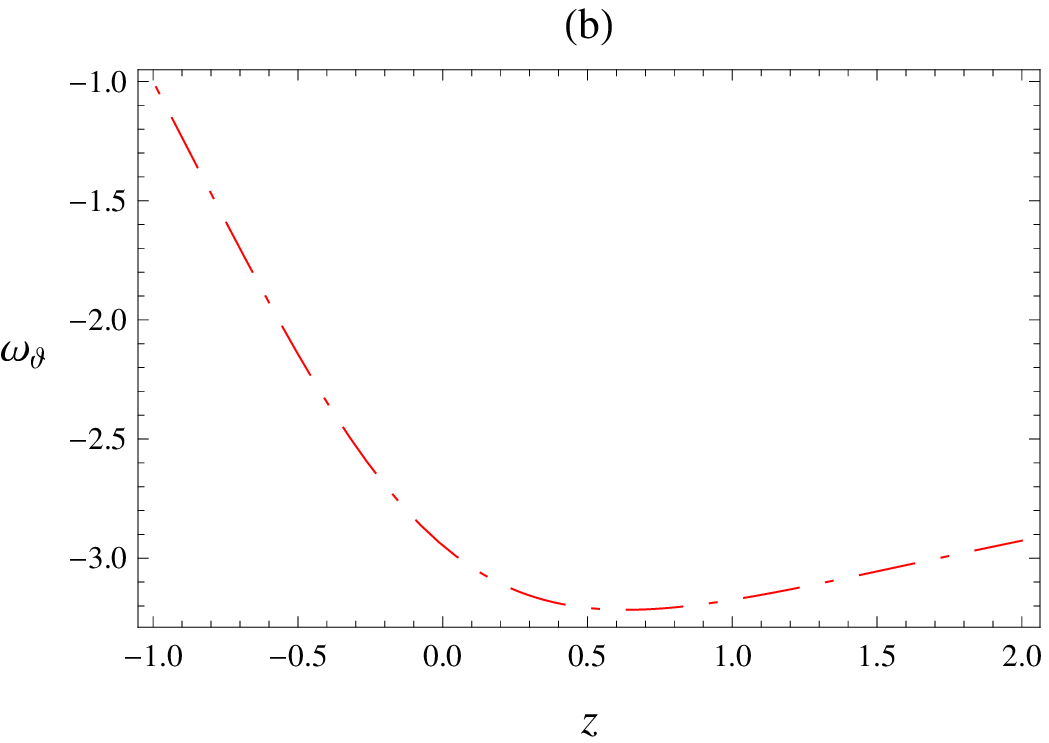, width=.495\linewidth,
height=2.2in} \caption{Evolution trajectories of $\Omega_\vartheta$ and $\omega_\vartheta$
for PDE with conformal time. We set $\mu=-1$ and $n=2$. }
\end{figure}
\begin{figure}
\centering \epsfig{file=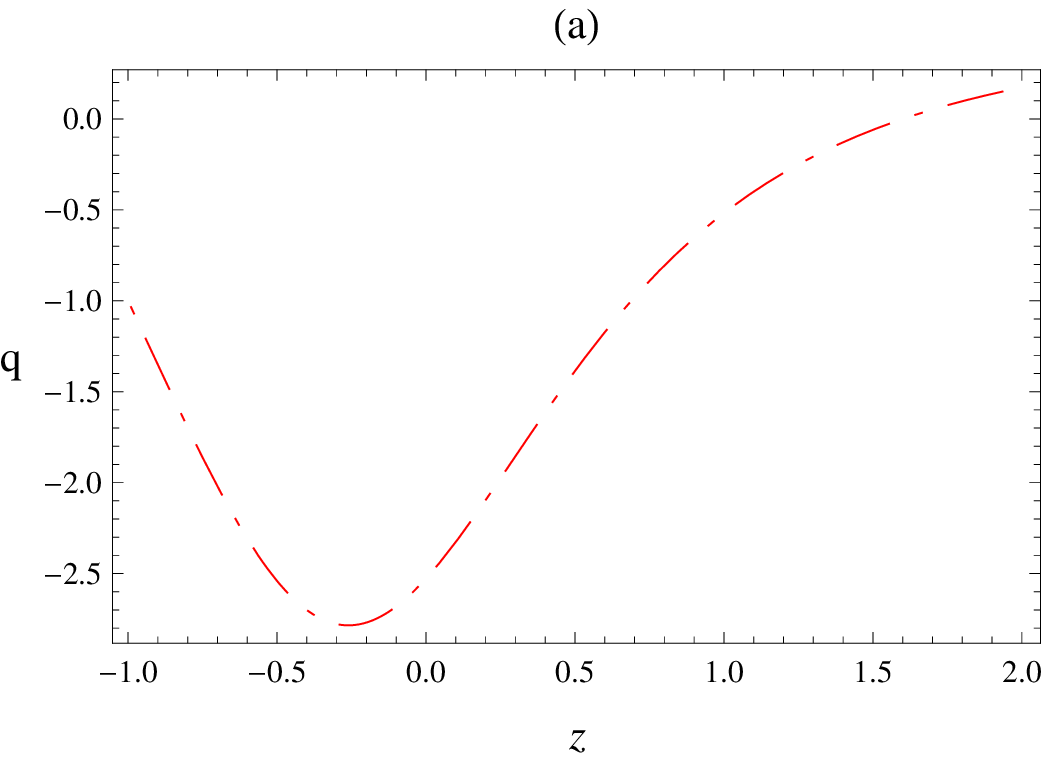, width=.495\linewidth,
height=2.2in} \epsfig{file=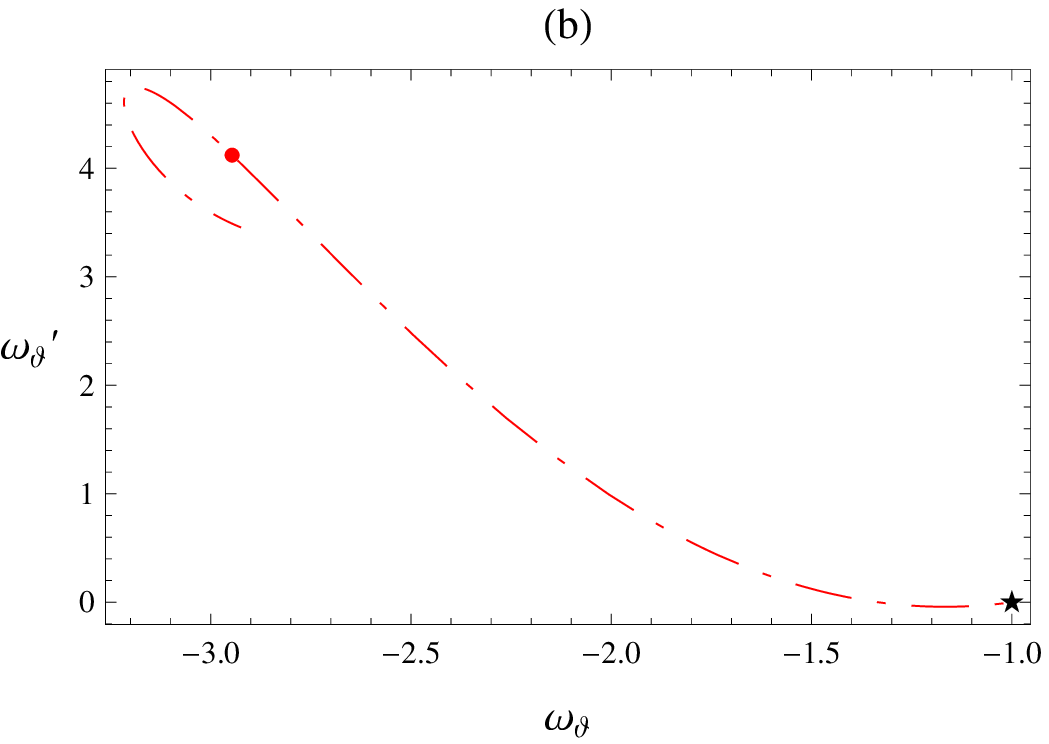, width=.495\linewidth,
height=2.2in} \caption{Evolution trajectories of $q$ and
$\omega_\vartheta-\omega'_\vartheta$ phase plane for $\mu=-1$ and
$n=2$. Star indicates the $\Lambda$CDM model with
$\omega_\vartheta=-1$ and $\omega'_\vartheta=0$.}
\end{figure}
\begin{figure}
\centering \epsfig{file=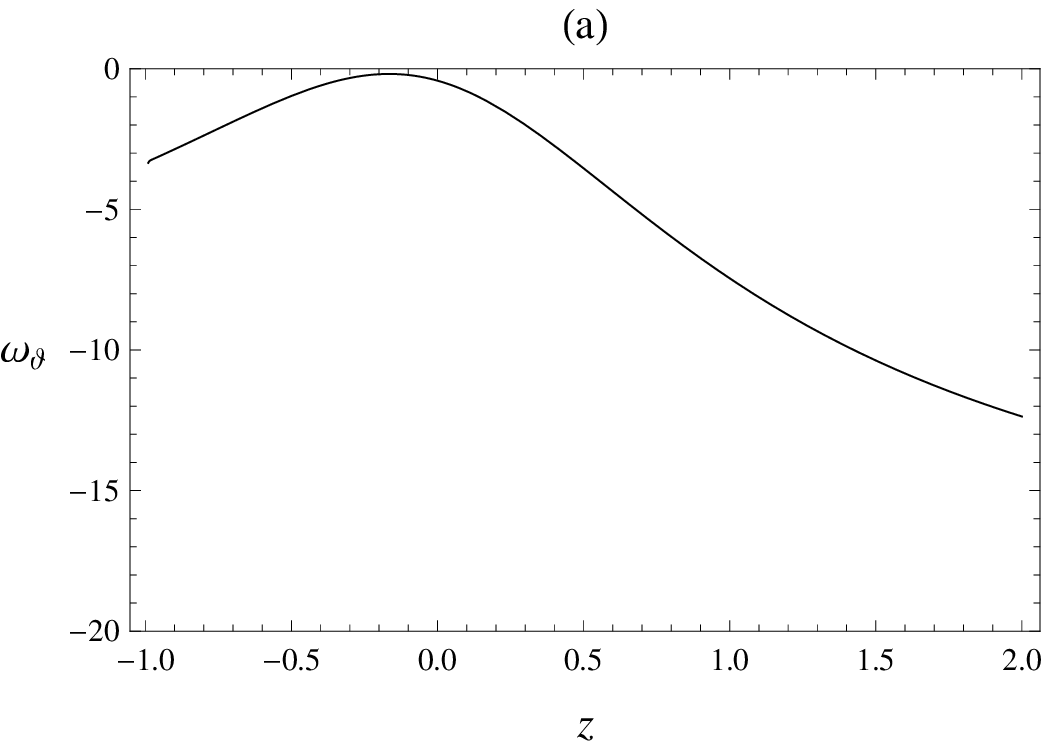, width=.495\linewidth,
height=2.1in} \epsfig{file=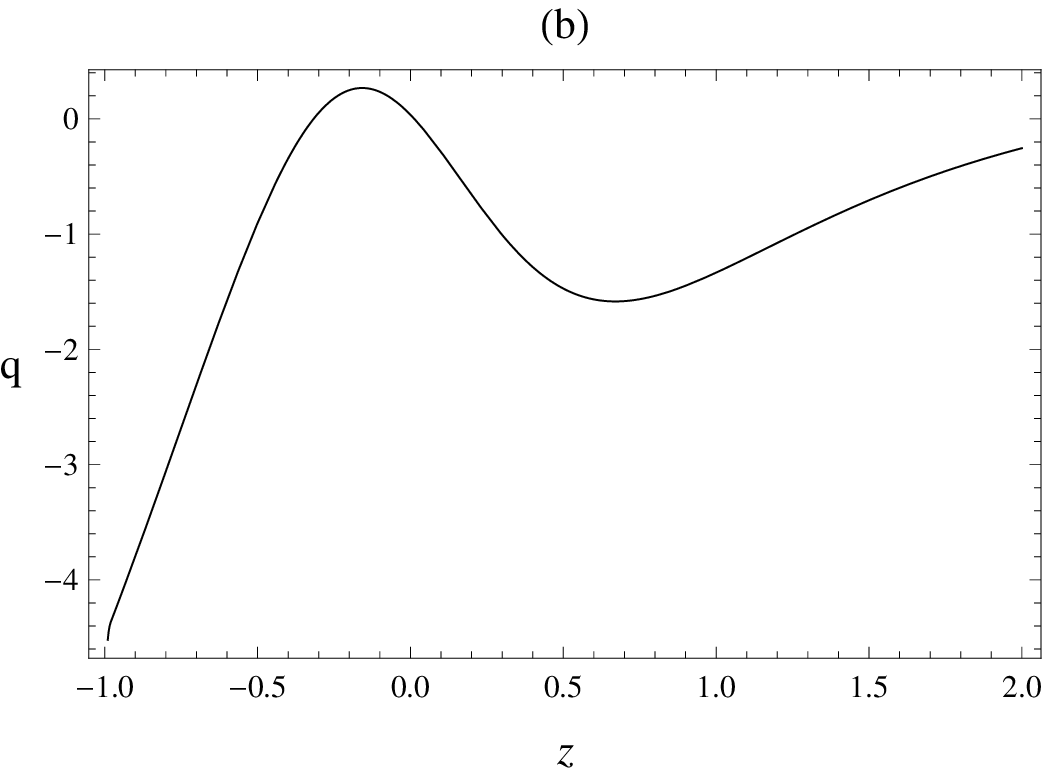, width=.495\linewidth,
height=2in} \caption{Evolution trajectories of $\omega_\vartheta$ and $q$
versus redshift for PDE with conformal time corresponding to $\mu=-55$ and $n=2$.}
\end{figure}

If we differentiate Eq.(\ref{20}) with respect to $x$, we have
\begin{equation}\label{21}
\omega'_\vartheta=-e^{-x}\left(1-\frac{3}{\mu}(1+\omega_\vartheta)\right)\left(1-\frac{\mu}{6}
-\frac{1}{3}\frac{d}{dx}\ln{\Omega_\vartheta}\right).
\end{equation}
Figure \textbf{9(b)} shows the evolution trajectories of $\omega'_\vartheta$
in $\omega_\vartheta-\omega'_\vartheta$ plane for PDE with cosmological time
scale. It is obvious that evolution in phase space of $\omega_\vartheta$ and
$\omega'_\vartheta$ results in $\Lambda$CDM model
$(\omega_{\vartheta}=-1,~\omega_{\vartheta}'=0$) as $z\rightarrow-1$ (or
$x\rightarrow{\infty}$). We analyze that if one sets $-55<{\mu}<-1$, it would
result in $\omega_\vartheta\rightarrow0$ (\emph{i.e.,} the matter dominated
universe) in future evolution. If one chooses $\mu\leqslant-55$, we can have
$\omega_\vartheta>-1$ in later times of cosmic evolution but in such case the
present day value is only consistent with WMAP$9$ observational results as it
represents the quintessence model of DE (see Figure \textbf{10}). For PDE
with cosmological time scale, we also present the evolution trajectories for
$\mu\geqslant3$ as shown in Figure \textbf{11}. It is found that for $\mu=3$,
we have $\Lambda$CDM regime whereas for $\mu\geqslant4$, it results in
quintessence era of the universe.
\begin{figure}
\centering \epsfig{file=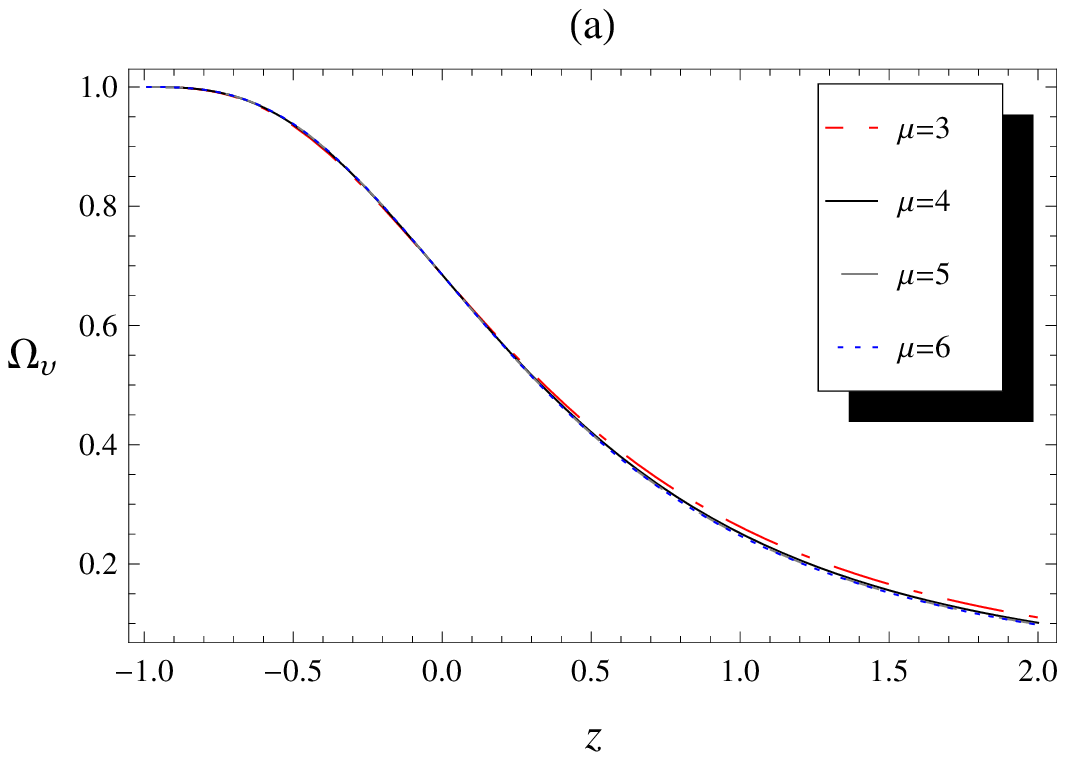, width=.495\linewidth,
height=2in} \epsfig{file=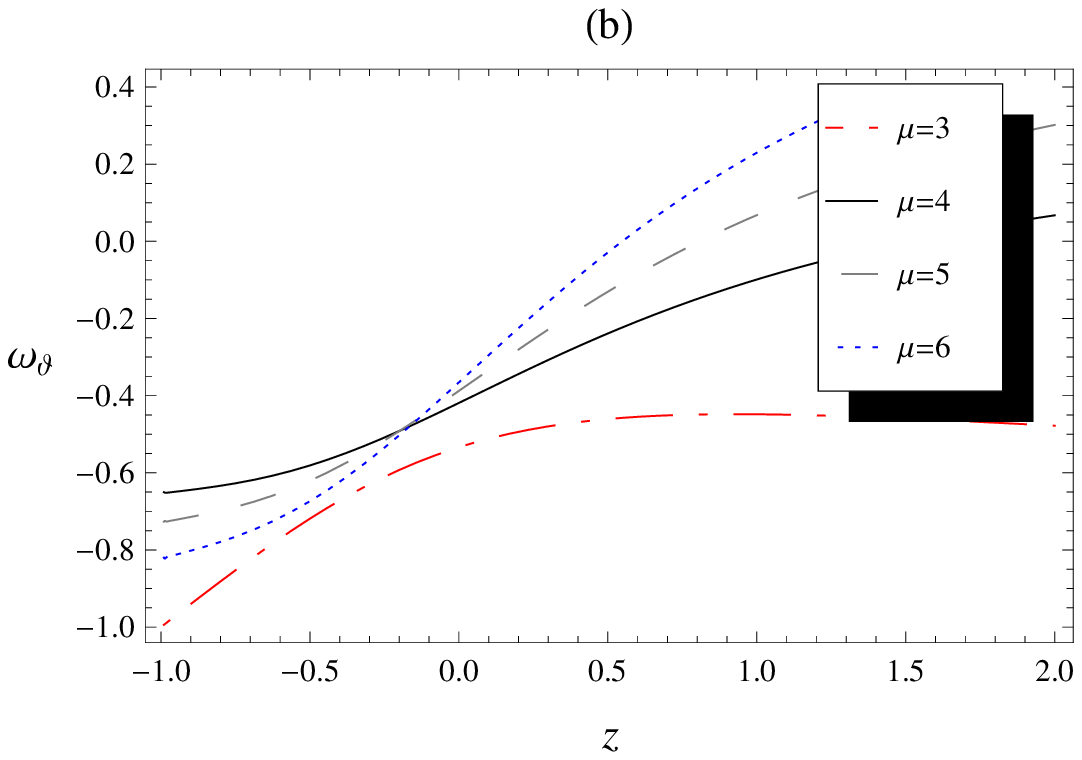, width=.495\linewidth,
height=2in} \epsfig{file=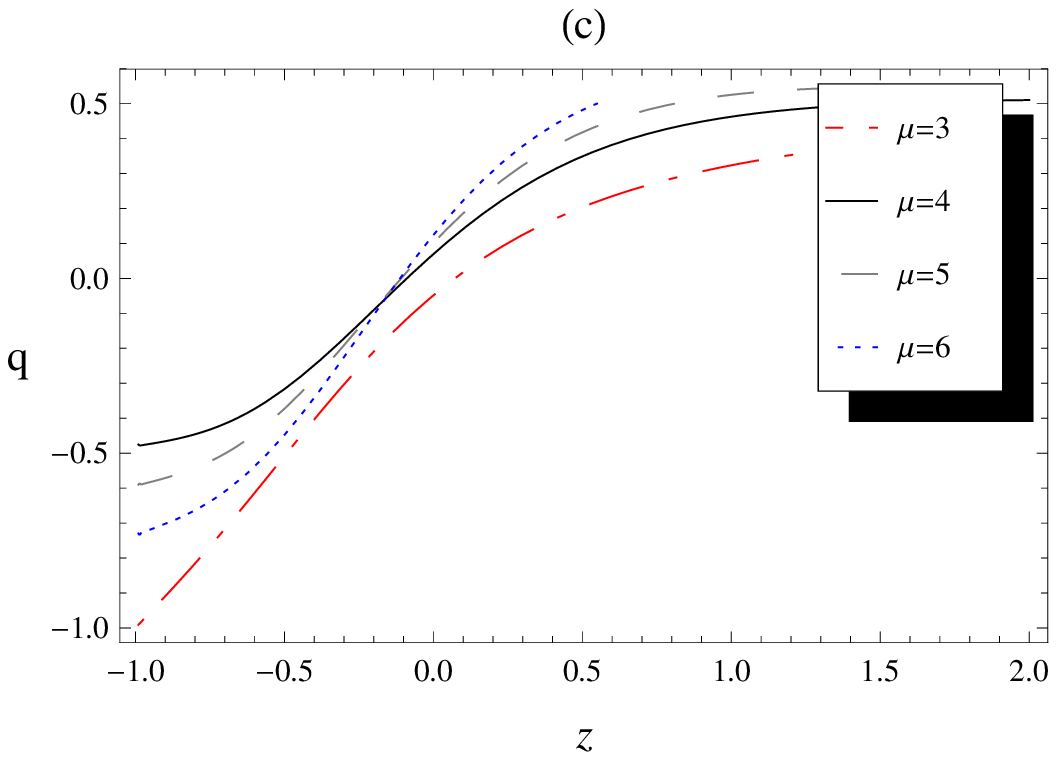, width=.495\linewidth,
height=2in} \epsfig{file=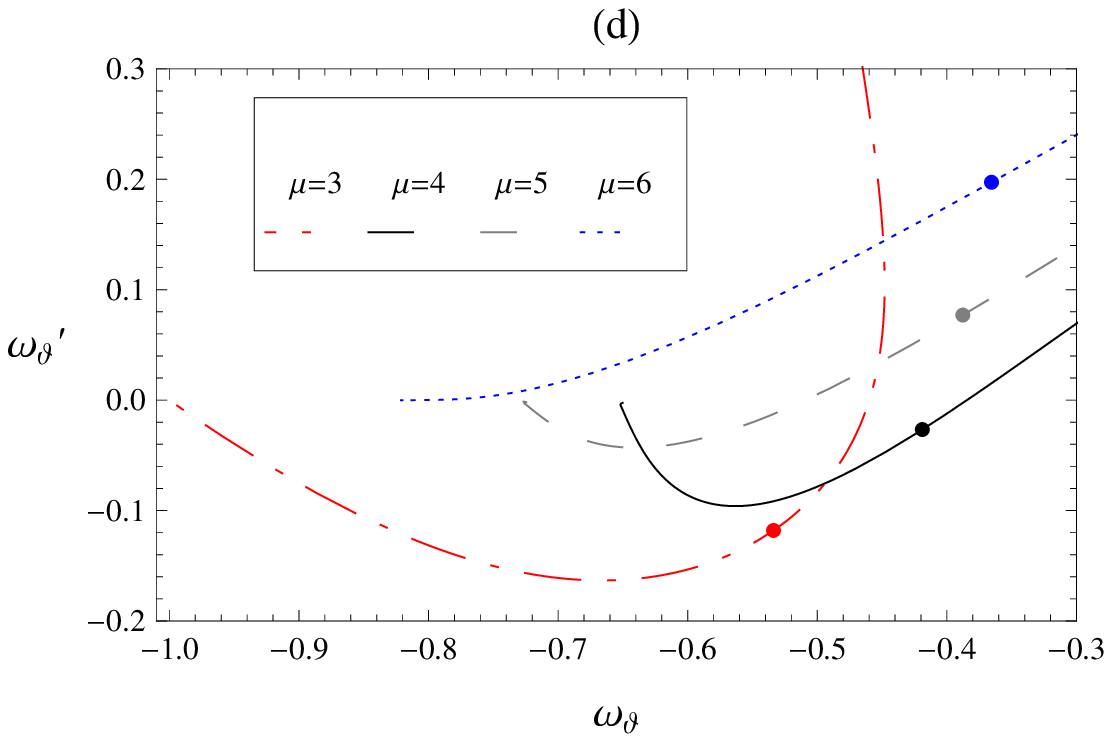, width=.495\linewidth,
height=2in}\epsfig{file=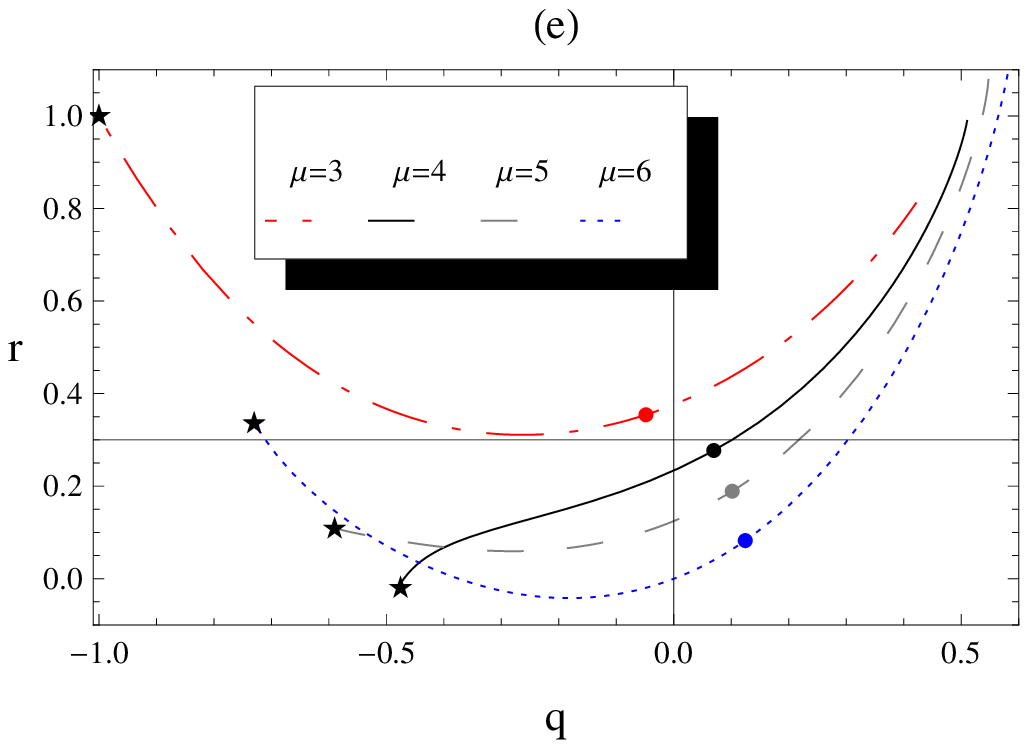, width=.495\linewidth,
height=2in}\caption{Evolution trajectories of
$\Omega_\vartheta,~\omega_\vartheta$ and $q$ versus $z$ and in
$\omega_\vartheta-\omega'_\vartheta$ and $q-r$ planes for PDE with
conformal time. Plot \textbf{(a)} shows the evolution trajectories
of $\Omega_\vartheta$ which results in DE dominated era in future
evolution. In plot \textbf{(b)} it is evident that for $\mu=3$, we
have de Sitter phase whereas $\mu\geqslant4$ represents the
quintessence regime. This behavior is well supported by the
evolution of $q$ in plot \textbf{(c)}. The evolution trajectories of
$\omega'_\vartheta$ are shown in plot \textbf{(d)} which indicate
the $\Lambda$CDM $(\omega_\vartheta ,\omega'_\vartheta)=(-1,0)$
model for $\mu=3$. Statefinder diagnosis can also be seen in plot
\textbf{(e)}.}
\end{figure}

\section{Conclusions}

In this paper, we study the phantom evolution of PDE with three cut-offs
namely particle horizon, event horizon and conformal age of the universe in
FRW spacetime. We explore these IR cut-offs to establish the consistent range
in PDE for parameter $\mu$ and also the phantom regime. Following (Li 2004),
the evolution equation of fractional energy density of DE $\Omega_\vartheta$
and dynamical relation of $\omega_\vartheta$ are formulated in these
settings. If $\mu=2$, one can determine the corresponding results in HDE with
particle horizon, event horizon and conformal age of the universe. We set the
present day values of parameters according to recent Planck observations and
present the evolution inconsistent with this data set.

Firstly, we have analyzed the non-interacting PDE with particle horizon and
shown the evolutionary paradigm of $\Omega_\vartheta$ and $\omega_\vartheta$.
If $\mu\geqslant3$, then we have purely matter dominated phase of the
universe since $\Omega_\vartheta\rightarrow0$ and $\omega_\vartheta>1$ for
$z\rightarrow-1$ as shown in Figure \textbf{1}. This choice is neglected in
search of some consistent models. For $\mu<0$, the plots of
$\Omega_\vartheta$ and $\omega_\vartheta$ are shown in Figure \textbf{2}
which represents that $\omega_\vartheta$ is in the phantom regime and never
intersects the phantom divide line in the whole cosmic history. This is
identical to that of PDE with Hubble horizon (Wei 2012). The evolution of
$\omega_\vartheta$ shows the phantom regime inconsistent with the current
observational results of Planck and WMAP$9$ data sets. We also plot
deceleration parameter and the phase space of $\omega_\vartheta$ and
$\omega'_\vartheta$ in Figure \textbf{3}. $\omega'_\vartheta$ lies in the
freezing region favoring the phantom evolution in this format of DE. The
statefinder diagnostic (Figure \textbf{4}) show that evolution trajectories
are consistent with phantom regime (Wu and Yu 2005, 2006). Thus, we conclude
that for realistic model of PDE with particle horizon one needs to set
$\mu<0$ and this choice is well supported by the results of Planck and
WMAP$9$ observations.

Secondly, we have explored PDE scenario in the light of event horizon. We are
mainly concerned with the choice of $\mu<0$ but for PDE with event horizon
one can also set $\mu=3$. For $\mu\leqslant-1$ and $\mu=3$, the evolution of
$\omega_\vartheta$ and $\Omega_\vartheta$ is shown in Figure \textbf{5}. It
is found that DE dominates in future evolution as
$\Omega_\vartheta\rightarrow1$ for $z$ approaching to $-1$. The EoS parameter
is in the quintessence regime in recent past which bisects the phantom divide
line and ends up in phantom era. For $\mu>2$, the acceptable results are
found only for $\mu=3$ whereas parameter $\mu>3$ does not show realistic
results. Figure \textbf{6(a)} favors the phantom DE showing sign flip of $q$
in recent past leading to $q<-1$. We also show the evolution trajectories in
$\omega_\vartheta-\omega'_\vartheta$ plane (Figure \textbf{6(b)}). It
represents the thawing region and in later times of the universe
$\omega'_\vartheta\rightarrow0$ with $\omega_\vartheta>-1$. These results are
also presented in $q-r$ plane (Figure \textbf{7}). It is found that for PDE
with event horizon the acceptable range of $\mu$ is $\mu\leqslant-1$ and
$\mu=3$.

Thirdly, we have used conformal time scale as IR cut-off for PDE. Initially,
we set $\mu=-1$ and plot the fractional density and EoS parameter of DE
(Figure \textbf{8}). Accordingly, $\Omega_\vartheta\rightarrow1$ shows the
dominance of DE and $\omega_\vartheta<-1$ in entire cosmic evolution which
approaches to $-1$ in ultimate fate of the universe. Consequently, cosmic
evolution ends up with cosmological constant regime avoiding the big rip
singularity which is identical to Wei (2012) for Hubble horizon. The
deceleration parameter $q$ shows the bouncing behavior of the universe where
the universe entered in accelerated expansion era in recent past and
concludes in de Sitter phase. These results are also favored by the phase
space of $\omega_\vartheta$ and $\omega'_\vartheta$ as shown in Figure
\textbf{9(b)}. It is found that the range $-55\leqslant\mu<-1$ results in
matter dominated cosmic evolution. If one sets
$\mu\leqslant-55,~\omega_\vartheta$ can show phantom evolution for later
times but in such case the present day value of $\omega_\vartheta$ is not
observationally consistent. Hence, for conformal age of the universe in PDE
scenario, the only acceptable value of $\mu$ is $-1$.

In (Sharif and Jawad 2013), authors discussed the PDE for non-interacting
case by defining EoS and other cosmographic parameters in terms of present
day values of $\Omega_\vartheta$, $\Omega_m$ and $H$. This study is confined
to present scenario and does not show the behavior in entire cosmic
evolution. Comparatively, the dynamical equation of $\Omega_\vartheta$ with
the initial condition $\Omega_{\vartheta0}=1-\Omega_{m0}$ implies that the
entire cosmic evolution of $\Omega_\vartheta$ as well as $\omega_\vartheta$
can be established. We remark that PDE with particle horizon leads to phantom
evolution if $\mu<0$ and for event horizon one can set both $\mu<0$ and
$\mu=3$. In case of conformal age of the universe, the result is quite
significant where only consistent value is $\mu=-1$ showing identical
behavior to that for the Hubble horizon (Wei 2012).

\vspace{0.25cm}

{\bf Acknowledgment}

\vspace{0.25cm}

We would like to thank the Higher Education Commission, Islamabad, Pakistan
for its financial support through the {\it Indigenous Ph.D. 5000 Fellowship
Program Batch-VII}.

\end{document}